\documentclass[12pt,a4paper]{article}

\usepackage{amsmath}
\usepackage{caption}
\usepackage{subcaption}
\usepackage{indent first}
\usepackage{jheppub}
\usepackage{pdfcomment}
\usepackage{tikz}
\usepackage{xspace}
\usetikzlibrary{shapes,matrix}

\newcommand{\bH}{\mathbb{H}}
\newcommand{\bC}{\mathbb{C}}
\newcommand{\bR}{\mathbb{R}}
\newcommand{\bZ}{\mathbb{Z}}
\def\Nequals#1{$\mathcal{N}{=}#1$}
\def\SU{\mathrm{SU}}

\def\USp{\mathrm{USp}}
\def\U{\mathrm{U}}
\def\SO{\mathrm{SO}}
\def\tr{\mathop{\mathrm{tr}}\nolimits}
\def\diag{\mathop{\mathrm{diag}}\nolimits}
\def\rank{\mathop{\mathrm{rank}}\nolimits}

\def\Tr{\mathop{\mathrm{Tr}}\nolimits}

\def\vev#1{\langle#1\rangle}

\def\beq#1\eeq{\begin{align}#1\end{align}}

\usepackage{xcolor}

\def\epf{very Higgsable\xspace}
\def\Epf{Very Higgsable\xspace}

\begin{document}

\title{6d \Nequals{(1,0)} theories on $T^2$ \\
and class S theories: part I}
\abstract{
We show that the \Nequals{(1,0)} superconformal theory on a single M5 brane on the ALE space of type $G=A_n, D_n, E_n$, when compactified on $T^2$, becomes a class S theory of type $G$ on a sphere with two full punctures and a simple puncture.  We study this relation from various viewpoints.  
Along the way, we develop a new method to study the 4d SCFT arising from the $T^2$ compactification of a class of  6d \Nequals{(1,0)} theories we call \emph{\epf.}
}
\author[1]{Kantaro Ohmori,}
\author[1]{Hiroyuki Shimizu,}
\author[1,2]{Yuji Tachikawa,}
\author[3]{and Kazuya Yonekura}
\affiliation[1]{Department of Physics, Faculty of Science, \\
 University of Tokyo,  Bunkyo-ku, Tokyo 133-0022, Japan}
\affiliation[2]{Institute for the Physics and Mathematics of the Universe, \\
 University of Tokyo,  Kashiwa, Chiba 277-8583, Japan}
\affiliation[3]{School of Natural Sciences, Institute for Advanced Study,\\
Princeton, NJ 08540, United States of America}
\preprint{IPMU-15-0028, UT-15-07}

\maketitle

\section{Introduction}\label{sec:introduction}

In the last few years, we learned a great deal about the class S theories, i.e.~the compactification of 6d \Nequals{(2,0)} theory on general Riemann surfaces with punctures. 
By starting from the 6d \Nequals{(2,0)} theories, which have a simple ADE classification, this construction gives a vast variety of 4d \Nequals2 theories, that come from the choice of the Riemann surfaces and punctures. 

There is another way to construct 4d \Nequals2 theories from 6d: namely, we can put 6d \Nequals{(1,0)} theories on $T^2$.  In this second method, there are no choice of the compactification manifold, but  there are a great number of \Nequals{(1,0)} theories in 6d as shown in a recent series of works \cite{Heckman:2013pva,Heckman:2015bfa,Bhardwaj:2015xxa}, thereby giving rise to a plethora  4d \Nequals2 theories.  A natural question, therefore, is how much overlap there is between these two constructions. 

\paragraph{Main objective.}
As a first step in this direction, in this paper we show that a small but natural subset of 6d \Nequals{(1,0)} theories on $T^2$ gives rise to a small but natural subset of class S theories.  Namely, we show that the 6d \Nequals{(1,0)} theory on a single M5-brane on the ALE space of type $G=A_n, D_n, E_n$, when compactified on $T^2$, becomes the class S theory of type $G$ on a sphere with two full punctures and a simple puncture.  
The 6d theories in question were called 6d $(G,G)$ minimal conformal matters in \cite{DelZotto:2014hpa}, and the 4d class S theories can be called the generalized bifundamental theories.  Using these terminologies, we can simply say that the $T^2$ compactification of the 6d minimal conformal matter gives  generalized bifundamental theory in 4d.

For $G=\SU(N)$ this relation is in a sense very trivial: a single M5-brane on the $\bC^2/\bZ_N$ singularity is just a bifundamental hypermultiplet of $\SU(N)^2$, and the class S theory of type $\SU(N)$ on a sphere with two full punctures and a single puncture is also a bifundamental \cite{Gaiotto:2009we,Gaiotto:2009hg}.
For $G=\SO(8)$, a single M5-brane on the $\bC^2/\Gamma_G$ singularity gives rise to the rank-1 E-string theory, as pointed out in \cite{Heckman:2013pva,DelZotto:2014hpa}.  The class S theory of type $\SO(8)$ on a sphere with two full punctures and a single puncture was studied in \cite{Chacaltana:2011ze}, and it was found that it gives the $E_8$ theory of Minahan and Nemeschansky. 
Therefore our objective is to show the relation in the other cases; but our analysis sheds new light even on the simplest of cases when $G=\SU(N)$.

\paragraph{Pieces of evidence.} In the rest of the paper, we  will  provide other pieces of evidences:
\begin{itemize}
\item In Sec.~\ref{sec:duality}, we follow the duality chain to show that the $T^2$ compactification of the 6d minimal conformal matter is a class S theory defined on a sphere with two full punctures and another puncture that cannot be directly identified with the present technology. 
\item In Sec.~\ref{sec:coulomb}, we compute and compare the dimension of the Coulomb branch both in 4d and in 6d.
\item In Sec.~\ref{sec:higgs}, we show that the Higgs branch of the 4d generalized fundamentals, when the $G^2$ flavor symmetry is weakly gauged, is given by the ALE space of type $G$. This is as expected from the 6d point of view.
\item In Sec.~\ref{sec:curve}, we compare the Seiberg-Witten curve of the 4d generalized bifundamental of type D and that of the 6d minimal conformal matter of type D in a certain corner of the moduli space and show the agreement. 
\item In Sec.~\ref{sec:anomaly}, we develop a method to compute the 4d anomaly polynomial of the compactification of a class of the 6d \Nequals{(1,0)} theories we call \emph{\epf}, apply that to 6d minimal conformal matters and show that they agree with the known central charges of 4d generalized bifundamentals.
\end{itemize}
We conclude with a short discussion in Sec.~\ref{sec:conclusions}.
These sections are largely independent of each other and can be read separately. 
In particular, the analysis given in Sec.~\ref{sec:anomaly} is quite general and applies to all 6d  theories we call \epf: these correspond, in the F-theoretic language of \cite{Heckman:2013pva,DelZotto:2014hpa,Heckman:2015bfa},  to theories whose configuration of curves $\mathcal C$ can be eliminated by a repeated blow-down of $-1$ curves.
Equivalently, the endpoint $\mathcal{C}_\text{end}$ is empty, and a further complex structure deformation makes the theory completely infared free without turning on any tensor vevs.
In other words,  the theory has a completely Higgsed branch where no tensor multiplet remains. This explains our terminology \emph{\epf}\footnote{The authors thank D. R. Morrison for the suggestion that led to this naming.}.

\section{Duality chain}\label{sec:duality}
Let us first try to follow the duality chain to show that the 6d minimal conformal matter on $T^2$ is a class S theory on a sphere with two full punctures and a simple puncture. We will see that there is one step we can not quite follow, due to our lack of knowledge of the 6d \Nequals{(2,0)} theory.

We start from a single M5-brane on the $\bC^2/\Gamma_G$ singularity. This gives a minimal conformal matter of type $G$ weakly coupled to $G^2$ gauge fields in 7d. By putting it on a torus, we should have a 4d theory with $G^2$ flavor symmetry, which is weakly coupled to $G^2$ gauge fields in 5d.

Let us say that the torus $T^2$ has complex structure $\tau$.  By compactifying on one side of $T^2$ and taking the T-dual of the other, we have Type IIB string theory on $\bR^{1,3}\times S^1\times \bR\times \bC^2/\Gamma_G$ with axiodilaton given by $\tau$, together with a single D3-brane filling $\bR^{1,3}$.   We now take the limit to isolate the low-energy degrees of freedom and ignore the center-of-mass mode of the D3-brane.  We have the 6d \Nequals{(2,0)} theory of type $G$ on $S^1\times \bR$, and the tension of the D3-brane becomes effectively infinite. Therefore we should have a BPS defect of codimension-2. With the class S technology currently available to us, we do not see how to directly identify this defect; let us call it $X$. 

We now take the limit where $S^1$ is small. Then we have a localized degrees of freedom, that is the class S theory of type $G$ on a sphere with two full punctures and a puncture $X$, coupled weakly to 5d $G$ gauge fields coming from the 6d \Nequals{(2,0)} theory of type $G$ on $S^1\times$ semi-infinite lines. 

Therefore we conclude that the 6d minimal conformal matter of type $G$, when compactified on $T^2_\tau$, is a class S theory of type $G$ on a sphere with two full punctures and a puncture $X$.   At present, the most we can say just using the duality chain is that we know that the puncture $X$ is the simple puncture when $G$ is either $\SU(N)$ or $\SO(8)$, and that the only statement that naturally generalizes this is that the puncture $X$ is always the simple puncture for arbitrary $G$. 

\section{Dimensions of the Coulomb branch}\label{sec:coulomb}
In this section, we compute the dimension of the Coulomb branch both in 4d and in 6d, and show that the results indeed agree. 

\subsection{6d perspective}
First, we take the 6d point of view. 
In Sec.~\ref{sec:duality} we followed the duality chain to map the 6d minimal conformal matter on $T^2$ to the Type IIB string on  $\bR^{1,3}\times S^1\times \bR\times \bC^2/\Gamma_G$ with axiodilaton given by $\tau$, together with a single D3-brane filling $\bR^{1,3}$.   
Instead of directly study the Coulomb branch in 4d, let us put the theory on another $S^1_R$ of radius $R$ and 
directly identify the hyperk\"ahler structure of the 3d Coulomb branch.
Take the T-dual of this $S^1_R$, and call it $\tilde S^1_{1/R}$. 
Then  lift the whole system back to M-theory. Here we are following the analysis of Appendix A.3 of \cite{Nekrasov:2012xe}.

We now have M-theory on $\bR^{1,2}\times \tilde S^1_{1/R} \times T^2_\tau \times \bR\times  \bC^2/\Gamma_G$ and a single M2-brane filling $\bR^{1,2}$.  The singularity has $G$ gauge multiplet on its singular loci, and the M2-brane can be absorbed into an instanton configuration. We conclude that the 3d Coulomb branch of the 6d minimal conformal matter on $S^1_R \times T^2_\tau$ is given by the one-instanton moduli space of gauge group $G$ on $\tilde S^1_{1/R}\times T^2_\tau\times \bR$. 

This gives an interesting new perspective on the tensor branch of the 6d minimal conformal matter. 
We consider an instanton configuration on $T^3\times \bR$.  By restricting the gauge field to $T^3$ at a constant ``time'' $t\in \bR$, we define the Chern-Simons invariant $CS(t)$. In our case, a single M5 gives a single M2 that becomes one instanton.  Let us say $CS(-\infty)=0$, then we have $CS(+\infty)=1$.

At $t=\pm\infty$, we need a zero-energy configuration, so the three holonomies $g_{1,2,3}$ around three edges of $T^3$ should commute.  We take them to be in the Cartan of $G$. By following the duality chain, we see that they can be identified with the original Wilson lines of $G^2$ used in the compactification. 
It is known that the Chern-Simons invariant of this flat gauge field on $T^3$ is 0 mod 1.
For simplicity, let us set $g_{1,2,3}=1$ at $t=\pm\infty$.

The quaternionic dimension of the moduli space including the center-of-mass motion  but with the holonomies at $t=\pm\infty$ fixed, is found by the Atiyah-Patodi-Singer index theorem~\cite{Atiyah:1975jf} to be\begin{equation}
d_{T^3,G}=h^\vee(G)-\rank(G)\label{eq:dim}
\end{equation} 
where $h^\vee(G)$ and $\rank(G)$ are the dual Coxeter number and the rank of $G$. The negative term is from the boundary contribution.\footnote{The theorem of \cite{Atiyah:1975jf} 
is valid if the gauge field approaches to the value at $t =\pm \infty$ exponentially rapidly.
That condition is satisfied by instanton configurations when the holonomies $g_{1,2,3}$ are generic so that the gauge group is broken to its Cartan.
Then the equation \eqref{eq:dim} follows from the fact that the 3d Dirac operator at $t =\pm\infty$  for the adjoint representation 
has $2\rank(G)$ zero modes and the $\eta$-invariant (excluding the zero modes)
of flat connections is zero.
By continuity, \eqref{eq:dim} should be valid even if we take $g_{1,2,3} \to 1$, although a direct analysis of this case is complicated.
}
Therefore, this is the dimension (plus one, due to the center-of-mass motion) of the Coulomb branch of the 4d theory we obtain by putting the 6d minimal conformal matter on $T^2$:
\begin{equation}
d_{T^2, (G,G)\ \text{min.~conf.~matter}} = h^{\vee}(G) -\text{rank}(G) -1 \label{eq:4ddim},
\end{equation}

Let us see these degrees of freedom in more detail below. These details can be skipped in a first reading. 

\paragraph{$G=\SU(N)$.}
When $G=\SU(N)$, $h^\vee(G)=N$ and $\rank(G)=N-1$, and then $d_{T^3,G}=1$. This corresponds to the fact that a single M5 on $\bC^2/\Gamma$ singularity only has the center-of-mass motion as the tensor branch degree of freedom.

\paragraph{$G=\SO(2N)$.}
Next, consider $G=D_N$. Recall~\cite{DelZotto:2014hpa} that a single M5-brane on $\bC^2/\Gamma_{D_N}$ singularity can split into two fractional M5-branes, and the emerging gauge group between the fractionated branes is $\USp(2N-8)$.  We should be able to identify this process in the 3d compactification. 
We have $d_{T^3,G}=N-2$, since  $h^\vee(D_N)=2N-2$ and $\rank(D_N)=N$. 
So we want to  identify these degrees of freedom in the instanton moduli space. 

First, recall that for $D_N=\mathrm{Spin}(2N)$ gauge group, there is a unique commuting triple $(g^*_1,g^*_2,g^*_3)$ that cannot be simultaneously conjugated into the Cartan; they can be chosen to be in a common $\mathrm{Spin}(7)$ subgroup, see Appendix  I of \cite{Witten:1997bs}.  The Chern-Simons invariant is $1/2$ mod 1 \cite{Borel:1999bx}, and the unbroken subgroup is $\mathfrak{so}(2N-7)$. 
Using this we  have the following one-instanton configuration on $T^3\times \bR$: 
\begin{itemize}
\item For $-\infty<t<t_0$, the configuration on $T^3$ is basically flat and given by $(g_1,g_2,g_3)=(1,1,1)$. $CS(t)$ stays almost constant close to 0.
\item At around $t=t_0$, the gauge configuration suddenly changes to $(g_1,g_2,g_3)=(g_1^*,g_2^*,g_3^*)$ 
dressed with holonomies in the Cartan of the commutant, $\mathfrak{so}(2N-7)$. 
$CS(t)$ jumps to 1/2.
\item   Again, for $t_0<t<t_1$, the configuration remains almost constant.
\item And then at around $t=t_1$, it suddenly changes back to $(g_1,g_2,g_3)=(1,1,1)$, making $CS(t)$ to jump to $1$. 
\end{itemize}
In these configurations, we see two parameters $t_{0,1}$ in addition to the $N-4$ holonomies from the Cartan $\mathfrak{so}(2N-7)$. In total, we have $N-2$. 

We can now identify the parameters $t_0$ and $t_1$ as the positions of the two fractional M5-branes, and the $\USp(2N-8)$ gauge group between the two fractionated M5-branes as the S-dual of $\mathfrak{so}(2N-7)$ we find here. The reason is that, after $T^3$ compactification, we have a 3d theory coupled to 4d \Nequals4 super Yang-Mills on the segment.  We know that the $S^1_R$ and $\tilde S^1_{1/R}$ are T-dual to each other, and therefore the coupling constants of the \Nequals4 super Yang-Mills in these two descriptions are inversely proportional to each other, and therefore the groups we see are related by S-duality. 

\paragraph{$G=E_n$.}
The analysis is completely similar to the cases above, using the data in \cite{Borel:1999bx}. For $G=E_6$, we have the following commuting triples: \begin{equation}
\begin{array}{r|ccccc}
\text{value $v$ of $CS$} & 0  & \frac13 & \frac12 & \frac23 \\
 \text{commutant $G_v$} & \mathfrak{e}_6 & \varnothing & \mathfrak{su}(3)  & \varnothing 
\end{array}.
\end{equation} Then the one-instanton configuration can go through these commuting triples. 
The degrees of freedom in the instanton moduli space are now the ``time'' of the jump from one commuting triple characterized by $CS=v_i$ to the next $CS=v_{i+1}$, together with the holonomies in the Cartan of $G_v$. 
In total, the equality \eqref{eq:dim} is reproduced if \begin{equation}
h^\vee(G)=\sum_{\text{possible value $v$ of $CS$}} (1+\rank G_v)\label{eq:eq}
\end{equation} and indeed this is satisfied. We also see that this is the sequence of gauge groups when the M5-brane gets fractionated on the $E_6$ singularity found in \cite{DelZotto:2014hpa}.

For $G=E_7$, the list of the commuting triples are \begin{equation}
\begin{array}{r|ccccccccc}
\text{value $v$ of $CS$} & 0  & \frac14& \frac13 & \frac12 & \frac23 & \frac34 \\
 \text{commutant $G_v$}& \mathfrak{e}_7 & \varnothing & \mathfrak{su}(2) &\mathfrak{usp}(6) & \mathfrak{su}(2)  & \varnothing 
\end{array}.
\end{equation} and for $G=E_8$, these are \begin{equation}
\begin{array}{r|cccccccccccc}
\text{value $v$ of $CS$} & 0  & \frac16 & \frac15&  \frac14& \frac13 & \frac25&  \frac12 & \frac35 & \frac23 & \frac34 & \frac45 & \frac56 \\
 \text{commutant $G_v$} & \mathfrak{e}_8 & \varnothing &\varnothing & \mathfrak{su}(2) &\mathfrak{g}_2 & \varnothing & \mathfrak{f}_4 & \varnothing & \mathfrak{g}_2 & \mathfrak{su}(2)  & \varnothing  & \varnothing
\end{array}.
\end{equation} In both cases, we can check that indeed the crucial equality \eqref{eq:eq} is satisfied, and the sequence of the groups are the S-dual of the ones that appear in the fractionation of the minimal conformal matter, see \cite{DelZotto:2014hpa}.

Actually, we can do a refined check of the above picture. Consider instanton configurations in which the gauge field at $t =-\infty$ ($t=+\infty$) is given by a commuting triple
with the commutant $G_i$ ($G_{i+1}$) and Chern-Simons number $v_i$ ($v_{i+1}$).
The dimension of the moduli space of these configurations is given by the Atiyah-Patodi-Singer theorem as
\begin{equation}
d_{i,i+1}=h^{\vee}(G)(v_{i+1}-v_{i})-\frac{1}{2}(\rank(G_i)+\rank(G_{i+1})),
\end{equation}
where $h^{\vee}(G)(v_{i+1}-v_{i})$ should properly be defined by the integration of the second Chern class in the adjoint representation. 
Using the above tables for the values of $v_i$ and the groups $G_i$, one can check (and it was indeed proved in \cite{Borel:1999bx}) that we always have $d_{i,i+1}=1$
for adjacent commuting triples in the tables.
This is interpreted as the fact that a fractional M5-brane has only the center-of-mass degrees of freedom.

Here, it is interesting to note that the equality \eqref{eq:eq} is exactly the one that guarantees the equality of the Witten index of pure \Nequals1 super Yang-Mills of gauge group $G$ computed both in the infrared using the gaugino condensation and in the ultraviolet using the semi-classical quantization. For more, see e.g.~\cite{Witten:2000nv}.

\subsection{Class S perspective}
Before moving to the class S theory side, let us recall the necessary notions of the nilpotent orbits. A nilpotent orbit for an nilpotent element $e$ in $\mathfrak{g}$ is the set of elements in $\mathfrak{g}$ that are $G_{\bC}$-conjugate to $e$. We denote the nilpotent orbit containing the nilpotent element $e$ by  $O_{e}$. 

There is a one-to-one correspondence between homomorphisms $\rho : \mathfrak{su}(2) \to \mathfrak{g}$, up to conjugacy, and nilpotent orbits $O_e$. The precise map is given by $e = \rho(\sigma^{+})$. For simplicity, we denote the nilpotent orbit containing $\rho(\sigma^{+})$ as $O_{\rho}$. 
In the case of $\mathfrak{g}=\mathfrak{su}(N)$, these homomorphisms are classified by Young diagrams as is well-known in the class S theory of type $A_{N-1}$.
In general, regular (and untwisted) punctures $X_i $ of the class S theory  of type $G$ are classified by these homomorphisms $\rho_i$.

One of the important ingredients in the relationship between the theory of nilpotent orbits and the class S theory is the Spaltenstein map $d$, defined for any simple Lie algebra $\mathfrak{g}$. This is a map
\begin{equation}
d: \{ \text{nilpotent orbits of} \: \mathfrak{g} \} \to \{ \text{nilpotent orbits of}\; \mathfrak{g}^{\vee} \},
\end{equation}
where $\mathfrak{g}^{\vee}$ is the Langlands dual of $\mathfrak{g}$. For example, when $\mathfrak{g}=\mathfrak{su}(N)$, this map is
to send a Young diagram to its dual diagram.
In this paper we only encounter the $\mathfrak{g}=\mathfrak{g}^{\vee}$ cases, so in the following we will assume this. Note that the Spaltenstein map is order-reversing, $d(O)\geq d(O')$ if $O\leq O'$ where the standard partial ordering for nilpotent orbits is defined so that $O_{e'}\geq O_{e}$ if $\bar{O}_{e'}\supset O_e$. 

The maximal orbit under this partial ordering is called the principal orbit $O_{\text{prin}}$ and is equal to $d(O_{0})$, the Spaltenstein dual to the zero orbit $O_{0}$. The dimension of the principal orbit is 
\begin{equation}
\text{dim}_{\bC} \, O_{\text{prin}} = \text{dim} (G) - \text{rank}(G) \label{eq:prin}.
\end{equation}
The next-to-maximal orbit is called the subregular orbit $O_{\text{subreg}}$ and is equal to $d(O_{\text{min}})$, where $O_{\text{min}}$ is the minimal nilpotent orbit.\footnote{This $O_{\text{min}}$ is defined by the homomorphism $\rho : \mathfrak{su}(2) \to \mathfrak{g}$ which
is used to embed the $\SU(2)$ one-instanton minimally into the group $G$. The dimension \eqref{eq:min}
is the same as the dimension of the one-instanton moduli space of $G$ minus the dimension of the center-of-mass of the instanton.} The dimension of the minimal orbit is 
\begin{equation}
\text{dim}_{\bC} \, O_{\text{min}} = 2(h^{\vee}(G)-1)\label{eq:min}.
\end{equation}

With these notions at hand, we put the $\mathcal{N}=(2,0)$ theory of type G on a Riemann surface of genus $g$ with regular and untwisted punctures $X_i $ which correspond to homomorphisms $\rho_{i}:\mathfrak{su}(2)\to \mathfrak{g}$. The complex dimension of the Coulomb branch of the resulting 4d \Nequals{2} theory is \cite{Chacaltana:2012zy}
\begin{equation}
d_{\text{class S}} = \sum_{i} d(\rho_{i}) + (g-1)\text{dim}(G), 
\end{equation}
where $d(\rho)$ is contribution from the punctures and is given by 
\begin{equation}
d(\rho) = \frac12 \text{dim}_{\bC} \, d(O_{\rho}).
\end{equation}

Let us apply this formula to the class S theory we are considering, namely, (2,0) theory of type G on a sphere with two full punctures and a simple puncture. The full puncture and the simple puncture are defined so that the corresponding nilpotent orbits are $O_{0}$ and $O_{\text{subreg}}$, respectively. Then, the Coulomb branch dimension is
\begin{align}
d_{\text{class S}} =& \text{dim}_{\bC} \, d(O_{0}) + \frac12 \text{dim}_{\bC} \, d(O_{\text{subreg}}) -\text{dim}(G) \nonumber \\
=& \text{dim}_{\bC} \, O_{\text{prin}} + \frac12 \text{dim}_{\bC} \, O_{\text{min}} -\text{dim}(G) \nonumber \\
=& h^{\vee}(G) - \text{rank}(G) -1,
\end{align}
where in the last line we used \eqref{eq:prin} and \eqref{eq:min}. This result agrees with \eqref{eq:4ddim}.

\section{Structure of the Higgs branch}\label{sec:higgs}
As the Higgs branch should remain identical under the $T^2$ compactification, the 6d theory and the 4d theory should have the same Higgs branch. We will check this below, at the level of complex manifolds.
It would be interesting to extend the analysis to the level of holomorphic symplectic varieties or hyperk\"ahler manifolds.

\paragraph{Type $\SU(N)$.} 
When the type $G$ of the theory we consider is $\SU(N)$, both the 6d minimal conformal matter and the 4d generalized bifundamental of type $\SU(N)$ are just a bifundamental hypermultiplet of $\SU(N)^2$. 
It naively seems there is not much to see here. However, we can still have some fun in this case, as we will see momentarily.  

Consider a single M5 brane on the $\bC^2/\bZ_N$ singularity. The 6d theory consists of the bifundamental of $\SU(N)^2$, weakly coupled to the 7d vector multiplet on the singular loci on the left and on the right of the M5 brane.  The Higgs branch of the system should describe the motion of the M5-brane on the $\bC^2/\bZ_N$ singularity. Therefore, we should be able to obtain $\bC^2/\bZ_N$ as the Higgs branch of the weakly-gauged bifundamental. Let us check this statement. In the 4d \Nequals1 notation, the bifundamental consists of $Q_i^a$ and $\tilde Q^i_a$, where $a,i=1,\ldots,N$. The invariant combinations under the $\SU(N)^2$ acting on the indices $a$ and $i$ are \begin{equation}
B=\det Q,\quad
\tilde B=\det \tilde Q,\quad
M = Q_i^a \tilde Q^i_a/N.
\end{equation} Note also that the bifundamental couples to the 7d gauge field via the moment maps \begin{equation}
\mu_i^j = Q_i^a \tilde Q^j_a - M \delta_i^j,\qquad
\tilde\mu^a_b = Q_i^a \tilde Q^i_b - M \delta^a_b.
\end{equation} They satisfy an important relation $\tr \mu^k = \tr \tilde \mu^k$ for any $k$.

The $\bC^2/\bZ_N$ singularity has $3(N-1)$ smoothing parameters, that can be naturally thought of as $(\mu_\bR, \mu_\bC)\in \mathfrak{su}(N)_\bR \times \mathfrak{su}(N)_\bC $, restricted to be in the Cartan; $\mu_\bR$ are the K\"ahler parameters for the resolution and $\mu_\bC$ the complex deformation. 
Therefore we can naturally identify this complex deformation parameter $\mu_\bC$ with $\mu\sim\tilde \mu$ above. 

Let us first consider the singular case $\mu_\bC=\mu=\tilde \mu=0$. Using the standard relation $\det Q_i^a \tilde Q^a_j  = \det Q \det \tilde Q= B\tilde B$ and $0=\mu^j_i = Q^a_i \tilde Q^j_a - M\delta^j_i$, we find \begin{equation}
B\tilde B= M^N.
\end{equation} This is indeed the equation of the $\bC^2/\bZ_N$  singularity. More generally, when \begin{equation}
\mu_\bC=\mu=\tilde \mu=\diag(m_1,\ldots, m_N),
\end{equation} we have $Q^a_i\tilde Q_a^j\sim \diag(m_1+M,\ldots,m_N+M)$. Therefore, we have \begin{equation}
B\tilde B=\prod_{i=1}^N (M+m_i),
\end{equation} which is again the equation of the deformed $\bC^2/\bZ_N$ singularity. 

\paragraph{General type.} Let us proceed to the general case. The 6d minimal conformal matter of type $G$, with the $G^2$ flavor symmetry weakly gauged, should have the Higgs branch of the form $\bC^2/\Gamma_G$, where $\Gamma_G$ is the finite subgroup of $\SU(2)$ of type $G$. Since the Higgs branch should be independent under the $T^2$ compactification, we should be able to check this using the class S description of the 4d generalized bifundamental. 

The Higgs branch of the class S theory in general was studied e.g.~in \cite{Moore:2011ee}. 
As discussed there, the Higgs branch of the class S theory of type $G$ on a sphere with two full punctures and a single regular puncture of arbitrary type is described as follows. We start from the Higgs branch $X_G$ of the $T_G$ theory, i.e.~the class S theory of type $G$ on a sphere with three full punctures. The hyperk\"ahler space $X_G$ has actions of $G^3$, and correspondingly has three holomorphic moment maps $\mu_1$, $\mu_2$, $\mu_3$ taking values in $\mathfrak{g}_\bC$.  
The hyperk\"ahler dimension of $X_G$ is \cite{Chacaltana:2012zy} \begin{equation}
\dim_\bH X_G= \rank G+\frac32(\dim G-\rank G).\label{dimXG}
\end{equation}

A  puncture is specified by a homomorphism \begin{equation}
\rho:\mathfrak{su}(2)\to \mathfrak{g}.
\end{equation} Such homomorphisms up to conjugation is known to be classified by the nilpotent element $e=\rho(\sigma^+)$ up to conjugation. Let $f=\rho(\sigma^-)$. We now define the Slodowy slice $S_e$ at $e$ by  \begin{equation}
S_e := \{ x+e  \mid [x,f]=0 \} \subset \mathfrak{g}_\bC.
\end{equation}  Then the class S theory of type $G$, on a sphere with two full punctures and a puncture specified by $e$, has the Higgs branch of the form \begin{equation}
Y_e=\mu_{1}^{-1} S_e
\end{equation} where we regarded $\mu_1$ as a map $X_G \to \mathfrak{g}_\bC$.

In our case we take $e$ to be the subregular element, since we want to have a simple puncture. 
The dimension is then \begin{equation}
\dim_\bH Y_e = \dim_\bH X_G - \dim_\bH O_\text{subreg} = \dim G + 1\label{dimY}
\end{equation}
where we used \eqref{dimXG} and \begin{equation}
\dim_\bC O_\text{subreg}= \dim G-\rank G -2.
\end{equation}

We would like to study the Higgs branch where the $G^2$ flavor symmetry is coupled to the $G_L \times G_R$ gauge multiplets in one higher dimension, associated to the $\bC^2/\Gamma_G$ locus from the left $(G_L)$ and the right $(G_R)$. 
Therefore the Higgs branch of the combined system is \begin{equation}
Z_e=Y_e  /\!/\!/ (G_L\times G_R).
\end{equation}
where $ /\!/\!/$ denotes the hyperk\"ahler quotient.
On a generic point of $Z_e$ , $G_L\times G_R$ is broken to its diagonal subgroup $G_\text{diag}$, 
since the $\bC^2/\Gamma_G$ locus is now connected and not separated by the M5 brane. 
The breaking from $G_L\times G_R$ to $G_\text{diag}$ should eat $\dim G$ hypermultiplets.
Subtracting this from \eqref{dimY}, we find that $\dim_\bH Z_e=1$: 
this agrees with our expectation that this Higgs branch describes the motion of an M5-brane along $\bC^2/\Gamma_G$ orbifold.
The question now is to see  that $Z_e=\bC^2/\Gamma_G$.

To see this, we use the following fact:
Let us say the $T_G$ theory has $G_1\times G_L\times G_R$ flavor symmetry,
and let us call the respective moment map operators  as $\mu_1$, $\mu_L$ and $\mu_R$.
Then 
 the Higgs branch operators of the $T_G$ theory, invariant under $G_L\times G_R$ 
 are just polynomials of $ \mu$   \cite{Maruyoshi:2013hja}. 
 We also know that $\mu$, $\mu_L$ and $\mu_R$ satisfy 
 the crucial relation \begin{equation}
\tr \mu_1^k=\tr \mu_R^k=\tr \mu_L^k \label{relation}
\end{equation} for any $k$.

Now consider the M5-brane on a singular $\bC^2/\Gamma_G$. 
This corresponds to the situation where the $G$ symmetry on the singular locus is unbroken.
This means that $\mu_L=\mu_R=0$,
which forces $\mu$ to be nilpotent via \eqref{relation}.  
Therefore the image of $\mu_1$ in $\mathfrak{g}_\bC$ is the variety $\mathcal{N}$ of nilpotent elements, 
and the final Higgs branch is therefore \begin{equation}
Z_e=S_e \cap \mathcal{N}.
\end{equation}  
The simple puncture corresponds to $e$ being the subregular element, and it is a classic mathematical fact by Brieskorn and Slodowy \cite{Brieskorn,Slodowy} that this space is the singularity $\bC^2/\Gamma_G$. \footnote{There are many mathematical ways to connect the simply-laced groups $G=A_n, D_n, E_n$, the finite subgroup $\Gamma_G$ of $\SU(2)$, and the singularity $\bC^2/\Gamma_G$. Probably the hyperk\"ahler quotient construction of Kronheimer \cite{Kronheimer} is more familiar to string theorists. But this result of Brieskorn and Slodowy was found much earlier in the mathematics literature.}

More generally, let us consider the case when the $\bC^2/\Gamma_G$ is deformed to a smooth manifold.
Such a smooth deformation can be parameterized by a generic element $h$ 
in the Cartan of $\mathfrak{g}_\bC$. 
The Higgs branch describing the motion of the M5-brane is then \begin{equation}
(O_{h,L} \times Y_e \times O_{h,R}) /\!/\!/ (G_L\times G_R) 
\end{equation}  
where  $O_{h,L}$ and $O_{h,R}$ are two copies of the orbit $\mathcal{O}_{h}$ of elements of $\mathfrak{g}_\bC$ conjugate to $h$,  and 
parameterize the vevs of the adjoint scalars in the 7d vector multiplets on the left and the right.\footnote{
More precisely, they arise as follows. To obtain supersymmetric configurations of the 7d gauge field, we have to solve Nahm's equations
on the half space $x_7>0$ ($x_7<0$) for $G_L$ ($G_R$) as in \cite{Gaiotto:2008sa}, where $x_7$ is the direction perpendicular to the M5-brane. 
The solution (at the complex structure level) is that a complex scalar $\Phi$ at $x_7=+0$ ($x_7=-0$) is in the orbit of $\Phi$ at $x_7=+\infty$ ($x_7=-\infty$).
These $\Phi(x_7 =\pm \infty)$ are just the vev of the field given by $\vev{\Phi}=h$. So the degrees of freedom from the 7d gauge field are
given by $\Phi( +0)$ and $\Phi(-0)$ which are in $O_{h}$. }
Since $G_L\times G_R$ is now broken to $\U(1)^{\rank G}$, the dimension of the resulting Higgs branch is \begin{equation}
2\dim_\bH O_h + \dim_\bH Y_e - (2\dim G-\rank G) = 1,
\end{equation} again the expected answer. 

Obtaining the Higgs branch itself is equally straightforward: we now have $\mu_1\in O_h$,
 and  the Higgs branch is now \begin{equation}
S_e \cap \mathcal{O}_{h}.
\end{equation} Again, it is a classic result of Brieskorn and Slodowy \cite{Brieskorn,Slodowy} that this space precisely gives the deformation of the $\bC^2/\Gamma_G$ singularity by the parameter $h$. 

\section{Seiberg-Witten curve}\label{sec:curve}

In this section, we compare the Seiberg-Witten curve of the 4d generalized bifundamental  and that of the 6d minimal conformal matter on $T^2$ when the type is $D_n$.  In principle we should be able to analyze the curves of  arbitrary type $G$ in a uniform fashion, but the authors have not been able to do that. 

The 6d conformal theory of type $D_N$, on the tensor branch, becomes $\USp(2N-8)$ theory with $2N$ flavors. Therefore, we should be able to reproduce the 4d curve of this theory by giving a suitable Coulomb branch vev to the 4d generalized bifundamental of type $D_N$.  

There is also another limit in which we can check the curve.
Instead of going to the 6d tensor branch, we can first reduce the 6d minimal conformal matter to 5d and add B-fields  to the ALE space. This makes the system one D4 brane on the $D_N$ orbifold, which is given by the quiver of the form
\begin{equation}
 \begin{tikzpicture}[scale=.9]
\node[draw,circle] (A) at (0,-.5) {1};
\node[draw,circle] (B) at (0,.5) {1};
\node[draw,circle] (C) at (1,0) {2};
\node[draw,circle] (D) at (2,0) {2};
\node[draw,circle] (E) at (3,0) {2};
\node[draw,circle] (F) at (4,-.5) {1};
\node[draw,circle] (G) at (4,.5) {1};
\draw (B)--(C)--(D)--(E)--(F);
\draw (A)--(C);
\draw (E)--(G);
\end{tikzpicture}\label{eq:quiverfig}
\end{equation} where a circle enclosing $i$ stands for an $\SU(i)$ gauge symmetry, and the edge between two gauge groups stands for the bifundamental. 
In the figure we used the case $N=6$ for explicitness. 
Adding B-fields corresponds to giving mass terms to the $D_N\times D_N$ flavor symmetries. 
Thus the 4d generalized  theory should also realize this quiver by the mass deformation.  

The relation of these two theories considered in 5d, namely the $\USp(2N-8)$ theory with $2N$ flavors and this $D$-type quiver theory, was called  ``a novel 5d duality'' in \cite{DelZotto:2014hpa}, and is the type $D$ version of the ``base-fiber duality'' of \cite{Bao:2011rc}. What we find here is that the corresponding 4d theories are both a deformation of a single class S theory, providing a 4d realization of these dualities. 

\paragraph{The curve of the generalized bifundamental.}
The generalized bifundamental of type $D_N$ is a class S theory of type $D_N$ on a sphere with two full punctures and a single simple puncture. Therefore, it has the following  Seiberg-Witten curve \begin{equation}
0=\lambda^{2N} + \phi_2(z) \lambda^{2N-2} + \cdots + \phi_{2N-2}(z) \lambda^2 + \phi_{2N}(z)
\end{equation}  where $\lambda=xdz/z$ is the Seiberg-Witten differential,
and $\phi_k(z)$ is a $k$-differential. We also need the single-valued-ness of $\tilde\phi_N(z)$ defined by $\phi_{2N}(z)=\tilde\phi_{N}(z)^2$. 

Let us put the full punctures at $z=0,\infty$ and the simple puncture at $z=1$. Writing $t=z-1$, the condition at the simple puncture is, according to \cite{Tachikawa:2009rb,Chacaltana:2011ze}\footnote{In terms of the Hitchin system, these rules are simply understood.
The Seiberg-Witten curve is $\det(\lambda - \Phi)=0$, where $\Phi$ is the adjoint field of the Hitchin system on the Riemann surface.
The condition $\phi_{2N}(z)=\tilde\phi_{N}(z)^2$ comes from $\det (-\Phi) = ({\rm Pfaff}(-\Phi))^2$ for $\mathfrak{so}(2N)$. The poles \eqref{eq:Dpoles} come from
$\Phi \sim e/t$, where $e$ is in the nilpotent orbit corresponding to the minimal embedding of $\mathfrak{su}(2)$ into $\mathfrak{so}(2N)$ as
$\mathfrak{su}(2) \subset \mathfrak{su}(2) \oplus \mathfrak{su}(2)=\mathfrak{so}(4) \subset \mathfrak{so}(2N)$.
In particular, one can check the relation of  the coefficients of $\phi_2$ and $\phi_4$ in \eqref{eq:Dpoles}.
}
\begin{equation}
\phi_2\sim \frac{2v_2}{t}dt, \quad
\phi_4\sim \frac{(v_2)^2}{t^2}dt , \quad 
\phi_{2k>4} \sim \frac{v_{2k}}{t^2}dt,\quad
\tilde\phi_N \sim \frac{\tilde v_N}{t} dt. \label{eq:Dpoles}
\end{equation}

From this we find that the curve is given by \begin{multline}
\frac{1}{z}\prod_{i=1}^{N}(x^2-m _i{}^2) + 2\underline{c} +  z\prod_{i=1}^{N}(x^2-\tilde m _i{}^2)  \\
=2 x^{2N} + M_2 x^{2N-2} + M_4 x^{2N-4} +  U_6 x^{2N-6}+U_8 x^{2n-2} + \cdots + U_{2N-2}x^2\label{foo}
\end{multline}
where $m_i$ and $\tilde m_i$ are mass parameters, $\underline{c}= \prod_i (-m_i {\tilde m}_i) $ so that $\phi_{2N}(z)=\tilde\phi_{N}(z)^2$ is satisfied,
$M_2$ and $M_4$ are quadratic and quartic polynomials of  $m_i$ and $\tilde m_i$ such that \eqref{eq:Dpoles} are satisfied for $\phi_2$ and $\phi_4$.
The Coulomb branch parameters are from $U_6$ to $U_{2N-2}$.

\paragraph{The $\USp$ theory.}
Let us next recall the curve of $\USp(2n)$ with $N_f+N_f'$ flavors: 
\begin{multline}
\frac{\Lambda^{2n+2-2N_f}}{z}\prod_{i=1}^{N_f}(x^2-m _i{}^2) + 2c + \Lambda^{2n+2-2N_f'} z\prod_{i=1}^{N_f'}(x^2-\tilde m _i{}^2)\\
=x^2(x^{2n}+u_2 x^{2n-2} + u_4 x^{2n-4} + \cdots + u_{2n})
\end{multline} where $c^2=\Lambda^{4n+4-2(N_f+N_f')} \prod_{i=1}^{N_f} (-m _i^2)\prod_{i=1}^{N_f'} (-\tilde m _i^2)$.
The differential is  $\lambda=xdz/z$. This curve in a hyperelliptic form was first found in \cite{Argyres:1995fw}. The form given above follows easily from the brane construction, see e.g.~\cite{Landsteiner:1997vd}.

Setting $2n=2N-8$, $N_f=N_f'=N$, the curve becomes \begin{multline}
\frac{1}{z}\prod_{i=1}^{N}(x^2-m _i{}^2) + 2\underline{c} +  z\prod_{i=1}^{N}(x^2-\tilde m _i{}^2)\\
=\Lambda^6 x^2(x^{2N-8}+u_2 x^{2N-10}  + \cdots + u_{2N-8})\label{bar}
\end{multline} where $\underline{c}=\Lambda^6 c$.

Coming back to the curve of the class S theory \eqref{foo}, we consider the regime $m_i, \tilde m_i \sim O(\epsilon)$, 
$\Lambda^6:= U_6 \sim O(1)$ and $U_k :=U_6 u_{k-6}\sim O(\epsilon^{k-6})$. 
Then the first three terms of  the right-hand-side of \eqref{foo} can be neglected\footnote{
This scaling limit is a little subtle due to the fact that our $\USp$ theory is not asymptotically free.
For example, we throw away the term $x^{2N}$ but retain both $z x^{2N}$ and $z^{-1} x^{2N}$.
}, 
and becomes \eqref{bar}. 
The identification of $U_6$ with some power of $\Lambda$ is natural since the vev of the tensor multiplet in 6d
is proportional to the gauge coupling of the $\USp(2N-8)$ gauge group.

This is consistent with the guess that this class S theory is the $T^2$ compactification of the  minimal conformal matter of type $D_N$. Also, we learn that the tensor branch scalar becomes $U_6$, of scaling dimension 6, independent of $N$, and is the coupling constant of the $\USp$ theory.

\paragraph{The $D$-type quiver.}
This is a completely different limit than the above $\USp$ limit.
Note first that the $D$-type quiver \eqref{eq:quiverfig} is in fact just the standard linear quiver with $\SU(2)^{N-3}$ gauge group, whose curve is well known.

We start from the curve  \eqref{foo} of the class S theory, and focus on the neighborhood of the simple puncture at $z=1$, 
by setting $z=(1+t)/(1-t)$, where $t$ is very small. The curve is  given, up to terms of $O(t^3)$, by\begin{equation}
\begin{array}{l@{}l@{}l@{}l@{}l@{}l@{}l}
0=t^2( x^{2N} & + c_2 x^{2N-2} & + c_4  x^{2N-4}& + c_6 x^{2N-6} & + \cdots &+ c_{2N-2}  x^2 & + c_{2N} )  \\
 & + 2t( \mu_2  x^{2N-2} & + \mu_4  x^{2N-4} &+ \mu_6  x^{2N-6} & + \cdots &+ \mu_{2N-2}  x^2 & + \mu_{2N} )   \\
 &  & + (\mu_2)^2  x^{2N-4} &+ U'_6  x^{2N-6} & + \cdots &  + U'_{2N-2}  x^2 & + b_{2N} 
\end{array}\label{iii}
\end{equation}
where we have defined
\begin{align}
&x^{2N} + c_2 x^{2N-2}+\cdots + c_{2N} =\frac{1}{2} \left(\prod_{i=1}^{N}(x^2-m _i{}^2)+\prod_{i=1}^{N}(x^2- \tilde{m} _i{}^2)  \right), \nonumber \\
&\mu_2 x^{2N-2}+\cdots + \mu_{2N} =-\frac{1}{4} \left(\prod_{i=1}^{N}(x^2-m _i{}^2)-\prod_{i=1}^{N}(x^2- \tilde{m} _i{}^2)  \right), \nonumber \\
& U'_k= - \frac{1}{4}U_k+c_k ,~~~~~b_{2N}= (-1)^{N}\frac{1}{4}\left( \prod_i m_i - \prod_i \tilde{m}_i \right)^2.   \nonumber
\end{align}
The differential is $\lambda=xdz/z \sim xdt \sim t dx$. 
One can check that the above curve is achieved in the scaling limit
$
t \sim O(\epsilon),~x \sim O(\epsilon^{-1}),~,m_i+\tilde{m}_i \sim O(\epsilon^{-1}), m_i - \tilde{m}_i =O(1),~U'_k \sim O(\epsilon^{-k+2})
$ and $\epsilon \to 0$. A similar limit was also considered in class S theories of type $A_{N-1}$ \cite{Hayashi:2014hfa}, and as in there,
the parameters $m_i-\tilde{m}_i$ may correspond to the masses of hypermultiplets in the quiver and $m_i+\tilde{m}_i$ may correspond to gauge couplings in 5d.

The coefficients of the terms $ t x^{2N-2}$ and $ x^{2N-4}$ are constrainted by the nonlinear relation of the pole coefficients at the simple puncture \eqref{eq:Dpoles}. 
This nonlinear relation is called a \emph{c-constraint} in \cite{Chacaltana:2011ze}.

Now, rewrite the curve as \begin{multline}
\underbrace{(\xi^{N} + c_2 \xi^{N-1} + \cdots + c_{2N})}_{=p(\xi)} \lambda^2 +2 \underbrace{(\mu_2 \xi^{N-1} +  \cdots +\mu_{2N})}_{=q(\xi)} (dx)\lambda \\ 
+ \underbrace{((\mu_2)^2 \xi^{N-2} + U'_6 \xi^{N-3} + \cdots+U'_{2N-2}x^2+ b_{2N})}_{=r(\xi)} (dx)^2 =0
\end{multline} where we introduced $\xi=x^2$. 
In the Seiberg-Witten curve of type $D$, $\pm x$ needs to be identified, and therefore this is  a natural choice.

Let us  put it in the Gaiotto form by  defining $\tilde\lambda=\lambda+q(\xi) dx/p(\xi)$, for which we have \begin{equation}
\tilde\lambda^2 + \varphi_2(\xi)=0.
\end{equation} We can check that $\varphi_2(\xi) =(d\xi)^2 (p(\xi)r(\xi)-q(\xi)^2)/4\xi p(\xi)^2$ has second-order poles at $N$ zeros of $p(\xi)$.
Thanks to the special forms of the coefficients of $tx^{2N-2}$ and $x^{2N-4}$  in \eqref{iii}, $\varphi_2(\xi)$ is finite at $\xi=\infty$.  
To study the behavior at $\xi=0$, recall the scaling limit described above. In that limit, we get $[  c_{2N}b_{2N} - (\mu_{2N})^2 ]/(\mu_{2N})^2 \to 0 $ and hence
the pole of $\varphi_2$ at $\xi=0$ disappears in the scaling limit. This is a consequence of the condition $\phi_{2N}(z)=\tilde\phi_{N}(z)^2$.
Therefore, we see that the curve is indeed that of the $\SU(2)^{N-3}$ quiver drawn above. 
With a little further effort, it can be checked that the residues of the double poles of $\varphi_2$ are proportional to $(m_i-\tilde{m}_i)^2$,
so $m_i - \tilde{m}_i$ are indeed proportional to the hypermultiplet masses of the quiver.

\section{\Epf theories and the central charges}\label{sec:anomaly}
In this section, we study the $T^2$ compactification of the class of 6d SCFTs that we call \epf.
We will determine the structure of the part of the Coulomb branch of the $T^2$ compactification that comes from the 6d tensor branch, and show in particular  that there is a point where one has a 4d SCFT. 
We will also show that the central charges $a$, $c$ and $k$ of the 4d SCFT can be written as a linear combination of the coefficients of the anomaly polynomial of the 6d SCFT. 
Since the 6d minimal conformal matters are \epf, we can apply the methods developed here to provide another check of our identification. 

Let us summarize the contents of this section. 
In Sec.~\ref{sec:str}, we introduce the class of the 6d SCFTs of our interest, namely the \epf theories.
In Sec.~\ref{sec:4d6drelation},
we recursively prove that \begin{itemize}
\item the $T^2$ compactification of a \epf theory gives  a 4d SCFT, and
\item the central charges of the resulting 4d SCFT can be written as a linear combination of coefficients of the anomaly polynomial of the 6d theory. 
\end{itemize}
In \ref{sec:centmatter}, we compute the central charges of the minimal conformal matter on $T^2$ by using the relationship with the anomaly polynomial of the minimal conformal matter.    
We will see that the resulting central charges indeed agree with the known central charges of the class S theory involved.

\subsection{\Epf theories}\label{sec:str}
Let us first define the class of 6d \epf theories.
In terms of the F-theoretic language of \cite{Heckman:2013pva,DelZotto:2014hpa,Heckman:2015bfa},  
a 6d SCFT can be characterized by the configuration $\mathcal C$ of curves on the complex two-dimensional base. 
We define a 6d SCFT to be \epf if successive, repeated blow-downs of $-1$ curves make $\mathcal C$ empty, or equivalently the endpoint ${\mathcal C}_\text{end}$ is empty. 
Then a further complex structure deformation removes the singularity completely.
In other words, there is a Higgs branch where the tensor multiplet degrees of freedom are completely eliminated, thus the word \epf. 
As examples,  the 6d $(G,G)$ minimal conformal matters and the general rank E-string theories are \epf, whereas the \Nequals{(2,0)} theory and the worldvolume theory of $Q >1$ M5 branes probing an ALE singularity are not \epf. 

We can also re-phrase the \epf condition without referring to the F-theory construction, in the following recursive fashion:
\begin{itemize}
\item Free hypermultiplets are \epf. 
\item An SCFT is \epf if 
	\begin{itemize}
	\item it has a one-dimensional subspace of the tensor branch on which the low-energy degrees of freedom consist of a single tensor multiplet, one or more \epf theories, possibly with a gauge multiplet $G$,
	\item such that the Chern-Simons coupling $S^{CS}$ of the self-dual two-form field of the tensor multiplet $B$,
		and its associated Green-Schwarz term $I^{GS}$ in the anomaly polynomial is\footnote{It may not be completely rigorous to write a
		Lagrangian like \eqref{eq:-1} for the self-dual 2-form $B$. But we will only need dimensional reduction of that 
		Chern-Simons term under the compactification on $T^2$ given by $2\pi \int b I_4$ where $b=\int_{T^2} B$.} 
		\begin{equation}
	S^{CS}=2\pi \int B \wedge I_4, ~~ I^{GS}_8 =\frac12 I_4^2, ~~ I_4 \supset \frac14 p_1(T)+\frac14 \Tr F_F^2-\frac14 \Tr F_G^2,\label{eq:-1}
	\end{equation} where the term $\Tr F_F^2/4$ is for the flavor symmetry, and the term $\Tr F_G^2/4$  is absent if there is no gauge multiplet.
	\end{itemize}
\end{itemize}

The condition \eqref{eq:-1} is a consequence of the fact that the tensor multiplet comes from a $-1$ curve, in the case of F-theoretic 6d SCFTs \cite{Sadov:1996zm,Ohmori:2014kda}.
Note that in our convention $\Tr F_G^2/4$ is the integrally normalized instanton density, and in particular the usual factor $(2\pi)^{-1}$ is absorbed into $F_G$.
Therefore, this means that the instanton-string has charge 1 under the tensor multiplet, which is the minimal consistent value under the Dirac quantization condition. The $p_1(T)$ etc.\ are the usual Pontryagin densities of the background metric.

We would like to study the $T^2$ compactification of a \epf theory.
Consider a tensor multiplet scalar $\phi$ associated to a $-1$ curve.
Classically, one of the 4d Coulomb moduli $u$ comes from the scalar  $\phi$, combined with the zero mode of the self-dual 2-form on the torus $b=\int_{T^2}B$:
\begin{equation}
u \sim \exp( \phi + 2\pi i b), \label{eq:coulombbranchoperator}
\end{equation}
where $b \simeq b+1$ due to the invariance under the large gauge transformation. 
The classical description in \eqref{eq:coulombbranchoperator} is valid in the region where $\phi$ is large compared to the size of $T^2$;  the moduli space can be significantly modified near $\phi \sim 0$. 

In general, the quantum corrections mix this variable $u$ with all the other Coulomb branch variables. 
However, in the case of the scalar $u$ for a $-1$ curve,
we can isolate a dimension-$1$ subspace ${\mathcal H}$ of the Coulomb branch parametrized by it.
This is because if a gauge multiplet is present on the minimally-charged tensor branch,
 the  4d gauge coupling of the gauge field  is infrared free,
 as we will prove below. 

Before proceeding, let us see two examples of this infrared freedom:
\begin{itemize}
\item First, the one-dimensional tensor branch of the $(D_k, D_k)$ minimal conformal matter for $k \ge 5$ supports the gauge group $\USp(2k-8)$. The number of flavors is $2k$, and therefore the system is infrared free as a 4d gauge theory. 
\item Second,  the F-theory realization of the $(E_6, E_6)$ minimal conformal matter has three curves with self-intersection $-1$, $-3$ and $-1$.   The middle $-3$ curve supports the gauge group $\SU(3)$. After the blowing down of the left and right $-1$ curves, the middle $-3$ curve becomes a $-1$ curve, and it gives a minimally-charged tensor multiplet.
This  still supports the $\SU(3)$ gauge group.
This gauge group is now coupled to two copies of the 4d version of rank-1 E-string theory, i.e.~the $E_8$ theory of Minahan and Nemeschansky.  One copy has the flavor current central charge $k_{E_8}/2=6$, and therefore two copies are worth $12$ flavors of $\SU(3)$ fundamentals.  Therefore the $\SU(3)$ gauge coupling is infrared free.
\end{itemize}

Thanks to the infrared freedom of $G$,  it is meaningful to talk about the origin of the Coulomb branch of $G$ even quantum mechanically. 
This determines the subspace $\mathcal H$.

\subsection{Structure of $\mathcal H$ and the central charges}\label{sec:4d6drelation}

\subsubsection{Properties to be recursively proved}
Now, we use the mathematical induction to prove the following properties of \epf theories: \begin{itemize}
\item The topology of $\mathcal H$ is always the same as that of the rank-1 E-string theory, namely, there are three singularities. Here, 
	\begin{itemize}
	\item two of them are the points where a single hypermultiplet becomes massless, and
	\item the third of them is a point at which the non-trivial SCFT appears, with the R-charge of the Coulomb branch operator $u$ being $12$. We call the resulting 4d SCFT as ${\mathcal T}_{4d}$.
	\end{itemize}
\item Writing the anomaly polynomial $I_8$  of the 6d theory ${\mathcal T}_{6d}$ as%
\footnote{Our normalizations and notations of 6d anomaly polynomials follows those in  \cite{Ohmori:2014kda}.}
\begin{align}
	I_8\supset \alpha p_1(T)^2+\beta p_1(T)c_2(R) +\gamma p_2(T) + \sum_i \kappa_i \, p_1 (T) \Tr F^2_i,
	\label{eq:coeffs}
\end{align}
 the central charges $a,c$ and flavor central charges $k_i$ of $i$-th flavor symmetry of the 4d theory ${\mathcal T}_{4d}$ are
\begin{align}
	a &= 24\alpha -12 \beta -18 \gamma, \nonumber\\
	c &= 64\alpha -12 \beta -8 \gamma, \nonumber\\
	k_i &= 192 \kappa_i.
	\label{eq:ack}
\end{align}
\end{itemize}

\subsubsection{Rough structure of the proof}
As the discussions will be rather intricate, here we provide the schematic structure of the inductive proof. 
The first step is to check the relations \eqref{eq:ack} for the free hypermultiplets.
In addition, we can check that free vector multiplets and free tensor multiplets both satisfy the relations \eqref{eq:ack}. 

The inductive step is to study the system of a minimally-charged tensor multiplet, with a \epf theory. 
There are two subcases: i) when there is no gauge multiplet, and ii) when there is a single gauge multiplet $G$. 
The subcase i) corresponds to the appearance of an E-string, for which the structure of $\mathcal H$ was studied long time ago \cite{Ganor:1996pc}.
In the subcase ii), the vev $u\in \mathcal H$ controls the dynamical scale $\Lambda(u)$ of the gauge group $G$. 
Since the coupling of $G$ is infrared free, $\Lambda(u)$ is the scale of the would-be Landau pole. 
From holomorphy, we expect at least one point on $u\in \mathcal H$ where $\Lambda(u)$ is zero. 
This is where we should have a nontrivial 4d SCFT ${\mathcal T}_{4d}$. 
From this, we will show that there will be two and only two additional singularities on $\mathcal H$, and that these are points where one massless hypermultiplet appears. 

In both subcases, we see that the structure of $\mathcal H$ is the same. 
Once this is known, we can employ the method of \cite{Shapere:2008zf}  to determine the central charges $a$, $c$ and $k$ of ${\mathcal T}_{4d}$ in terms of the 6d anomaly polynomial. This then confirms the general relation \eqref{eq:ack}, completing the inductive process.

\subsubsection{Structure of $\mathcal H$}
Now let us start the full discussion of the inductive step.
We first would like to establish the singularity structure of $\mathcal H$. 
When there is no gauge multiplet on the tensor branch, we have the E-string theory, for which the structure of $\mathcal H$ is known \cite{Ganor:1996pc}. 
There is a point where we have a 4d $E_8$ theory of Minahan and Nemeschansky, 
where the R-charge of the Coulomb branch operator $u$ is 12 and therefore the scaling dimension is 6. 
This is true for higher-rank E-string theory too. 

Let us next consider the case with a gauge multiplet with gauge group $G$. 
Denote the \epf theory on this tensor branch by $\mathcal S$. 
The low energy theory on this branch  consists of $\mathcal{S}$, the non-abelian gauge multiplet $G$, and a $\U(1)$ (or tensor) multiplet containing $u$,
and we want to show that there is a point at which they are combined into a single strongly interacting superconformal theory $\mathcal{T}$.

The theory $\mathcal{S}$ has flavor symmetry $H$ (not necessarily simple), and its subgroup $G \subset H$ is gauged by the non-abelian gauge group.
The commutant $F$ of $G$ in $H$ is the flavor symmetry of the total system.
The term proportional to $\Tr F_G^2 p_1(T)$ in the total 6d anomaly polynomial is given by  \begin{equation}
I^{\mathcal S} + I^\text{tensor} + I^\text{gaugino} + I^{GS}
\supset 
 (\kappa_{G}^{{\mathcal S}_{6d}}  - \frac{h^{\vee}_G}{48} -\frac1{16}) \Tr F_G^2  p_1(T).
\end{equation}
The gauge group $G$ is anomaly free in 6d, therefore \begin{equation}
48 \kappa^{\mathcal{S}_{6d} }_{G}-h^\vee_G=3.
\end{equation}
Using the inductive assumption \eqref{eq:ack}, we see that \begin{equation}
k_G^{\mathcal S_{4d}} - 4h^\vee_G =  12 > 0\label{eq:anomfoo}
\end{equation} which means that the one-loop beta function is positive and the $G$ gauge coupling in the 4d theory is infrared free. 
This guarantees that we can isolate the subspace $\mathcal H$ as we repeatedly emphasized above.

In addition, away from the singularities on $\mathcal H$, we can safely introduce the exponentiated complexified coupling \begin{equation}
\eta(u):= \Lambda_0^{-6}e^{2\pi i\tau_G(u)}
\end{equation}  of the 4d $G$ gauge field, defined at an arbitrary (but sufficiently small) renormalization group scale $\Lambda_0$, where $-6=2h^\vee-k_G^{\mathcal S_{4d}}/2 $
is the coefficient of the one-loop beta function. 
The Green-Schwarz coupling \eqref{eq:-1} in 6d gives the 4d coupling 
\beq
-2\pi b \cdot \frac{1}{4} \Tr F_G^2 \label{eq:NGanom}
\eeq
after the compactification, at least for large values of $\phi$. 
Then $-2\pi b$ can be identified as the theta angle of the gauge group in 4d, ${\rm Re}(\tau_G) = -b$.
Together with the definition \eqref{eq:coulombbranchoperator} of $u$ and holomorphy of $\tau_G(u)$, 
we can see that in the region $|u|\to \infty$, the $G$ coupling behaves as 
$\eta(u)\sim u^{-1}$. 
We expect $\eta(u)$ to be a single-valued meromorphic function on $\mathcal H$.\footnote{
In the case in which $\eta(u)$ can have multivalued behavior, there must be a duality transformation relating those multi-values of the coupling constant.
For example, in Seiberg-Witten theory of a massless $\U(1)$ field, a free $\U(1)$ has an electric-magnetic dual description 
which changes the coupling as $\tau \to -1/\tau$, and this was crucial for the multivalued behavior of $\tau$~\cite{Seiberg:1994rs}. 
However, in our case, $\eta(u)$ is properly understood as the position of the Landau pole
of the infrared free gauge field, and in particular it is a dimensionful parameter. There seems to be no duality transformation which 
sends one value of the Landau pole to another, and hence $\eta(u)$ is single-valued. 
However, if the gauge group were conformal rather than infrared free, we could have
multivalued coupling constant on the moduli space due to S-duality of the conformal gauge group. 
Such a situation indeed appears in other theories and will be discussed elsewhere.
}
We do not expect any zeroes of $\eta(u)$: if there is a zero, the gauge coupling of $G$ becomes extremely weakly coupled there, but we do not know of any physics to explain it. 
A single valued meromorphic function with the asymptotic behavior $\eta(u)\sim u^{-1}$ must have just a single simple pole. 
We define  the coordinate origin of $\mathcal H$ so that the pole of $\eta(u)$ is at $u=0$. This is the strongly interacting point where $\mathcal{T}_{4d}$ appears.
The 4d theta angle of the $G$ gauge multiplet at $u \neq 0$ is just given by the phase of $u$, globally on $\mathcal H$.

Slightly away from this point $u=0$, the infrared physics is the theory $\mathcal S_{4d}$ coupled to the $G$ gauge multiplet. The $\U(1)_RG^2$ is anomalous by the amount \eqref{eq:anomfoo}.
At the SCFT point this $\U(1)_R$ symmetry should be restored and it must be anomaly free. By the anomaly matching, 
the Nambu-Goldstone boson of the spontaneously broken $\U(1)_R$ at $u \neq 0$ must contribute $-12$ to the anomaly $\U(1)_RG^2$ via the coupling 
\eqref{eq:NGanom}, where $2\pi b$ should be interpreted as the phase of $u$ in the small $u$ region.
This can be done by assigning the $\U(1)_R$ charge 
\begin{equation}
R[u]=12\label{eq:toRcharges2}
\end{equation} to the $u$ near $u=0$. Then the total $\U(1)_RG^2$ anomaly is cancelled.

We now show that there are two more singularities on $\mathcal H$ and that these two points are associated with an additional massless hypermultiplet. The proof goes as follows: consider the Seiberg-Witten curve on $\mathcal H$ given by 
\begin{equation}
y^2 = x^3 + f(u) x + g(u).\label{eq:swcurve}
\end{equation}
This curve is for describing the effective action of the $\U(1)$ gauge field coming from the 6d tensor multiplet, and it should not be confused with $\tau_G$
which is the coupling of the non-abelian gauge group $G$.

Using the special coordinate on $\mathcal H$ related to the curve \eqref{eq:swcurve} via
\begin{equation}
\frac{da}{du} = \int_A \frac{dx}{y}, \hspace{0.5cm} \frac{da_D}{du} = \int_B \frac{dx}{y},\label{eq:special}
\end{equation}
where $A$ and $B$ are the two independent cycles of the torus, the metric on $\mathcal H$ is 
\begin{equation}
ds^2 = \text{Im}\, (da^* da_D) = \text{Im}\, \biggl{(}\frac{da}{du}^* \frac{da_D}{du}\biggr{)} |du|^2.\label{eq:metric}
\end{equation}
The complex structure $\tau = da_D/da$ is constant at $|u| \to \infty$ since it is given by the complex structure of the $T^2$ used in the compactification 
from 6d to 4d. 
Then $f$ and $g$ should behave as $f \sim u^{4n}$ and $g \sim u^{6n}$ for some $n$ for large $u$. Furthermore, the metric on $\mathcal H$ at $|u| \to \infty$ is the cylindrical one $ds^2  \sim d\phi^2+(2\pi db)^2 \sim |d \text{log}\, (u)|^2$ since it just comes from the compactification of a free tensor multiplet.
Substituting the asymptotic behavior of $f$ and $g$ to \eqref{eq:swcurve}, \eqref{eq:special} and \eqref{eq:metric}, we obtain $n=1$.

Next, let us consider the singularity at $u=0$. We set the asymptotic behavior of $f$ and $g$ at $u=0$ as $f \sim u^{4p}$ and $g \sim u^{6q}$. Then, the R-charge of $x$ and $y$ in \eqref{eq:swcurve} is 
\begin{equation}
R[x] = 2r R[u],  \hspace{0.5cm} R[y]= 3r R[u],\label{eq:xyRcharge}
\end{equation}
where $r = \text{min}\, (p,q)$. The R-charge of the Seiberg-Witten differential $\lambda$, which is the same as the R-charge of  
$u(\partial \lambda/ \partial u) = u dx/y$, is fixed to $2$ since its scaling dimension is 1.
Using \eqref{eq:xyRcharge}, the relation
\begin{equation}
(1-r) R[u] = 2
\end{equation}
holds. The fact $R[u] =12$ at $u=0$ leads to $r=q=5/6$ and then $p(>r)$ is $1$. Thus we obtain $f \sim u^4$ and $g \sim u^5$ near $u \sim 0$. Therefore the behavior of $f$ and $g$ on $\mathcal H$ is
\begin{equation}
f \sim u^4, \hspace{0.5cm} g \sim u^5 + u^6. \label{eq:E1curve}
\end{equation}
In particular, examining the discriminant $\Delta = 27 f^3 + 4 g^2$, there are two more singularities other than $u=0$ and that they are massless hypermultiplet points.

\subsubsection{Central charges from measure factors}
Before proceeding, let us very briefly recall the method of \cite{Shapere:2008zf} to compute the central charges $a$, $c$ and $k$ of 4d \Nequals{2} SCFTs from their topologically twisted cousins; we almost follow the conventions used in that paper. 
We put an \Nequals{2} supersymmetric field theory in 4d on a curved manifold with a non-trivial metric and a background gauge field for the flavor symmetry $F$ via the twisting of the $\SU(2)_R$ R-symmetry with one of the $\SU(2)$'s of $\SU(2) \times \SU(2) \simeq \SO(4)$ of the tangent bundle. 
In the following we assume that $F$ is nonabelian. We denote the Euler characteristic of the 4-manifold by $\chi$, the signature by $\sigma$ and the anti-instanton number for $F$ by $n$. We also denote by $u$ a set of gauge and monodromy invariant coordinates on the Coumlomb branch.  

The path integral of the twisted theory is given as follows
\begin{equation}
Z = \int [du][dq] A^{\chi}(u)B^{\sigma}(u) C^{n}(u) \text{exp}(-S_{\text{low energy}}).\label{eq:twistedpathintegral}
\end{equation}
Here $[du]$ and $[dq]$ are the path integral measures for the massless vector multiplets and other massless multiplets on the generic point of the Coulomb branch. The $A(u)$, $B(u)$ and $C(u)$ are factors induced by the non-minimal coupling of $u$ to the non-trivial background
which are given, up to coefficients, as $ \int \log A(u)  \tr R \wedge \tilde{R}$, $\int \log B(u)  \tr R \wedge R$ and 
$\int \log C(u) \tr F_F \wedge F_F$ in the effective action on the Coulomb branch. 
Supersymmetry requires that they are holomorphic. See \cite{Witten:1995gf} for details.

On a singular point on the Coulomb branch, we can have nontrivial superconformal field theory.
Then there must be an enhanced $\U(1)_R$ symmetry at each of these points, although $\U(1)_R$ need not be defined globally on the
Coulomb moduli space. 
The coefficients of the anomaly of $\U(1)_R$ under background fields are related to the central charges $a,c,k$ by supersymmetry as
\beq
\int d^4 x \partial_{\mu} j^\mu_{\U(1)_R} = (4a-2c)\chi+3c \sigma+kn,\label{eq:U1Ranom}
\eeq
where the term $\chi$ is due to twisting $\SU(2)_R$.
By using the same anomaly matching which was used to derive \eqref{eq:toRcharges2}, 
the central charges $a$, $c$ and $k$ are obtained as \cite{Shapere:2008zf}
\begin{align}
a &= \frac14 R[A] + \frac16 R[B] + a_\text{generic},  \label{eq:centralchargea} \\
c &= \frac13 R[B] + c_\text{generic}, \label{eq:centralchargec} \\
k &=  R[C]+k_\text{generic} \label{centralchargek}
\end{align} where $R[A,B,C]$ are the $\U(1)_R$-charges of the measure factors $A(u),B(u),C(u)$,
and $(a,c,k)_\text{generic}$ are the central charges at a generic point on the Coulomb branch.
The terms proportional to $R[A,B,C]$ are the contributions from $\U(1)_R$ Nambu-Goldstone bosons near each superconformal point.
For the gauge group $G$, what we have found in the previous subsection may be rephrased as
$k|_G=0$, $k_\text{generic}|_{G}=k_G^{\mathcal S_{4d}} - 4h^\vee_G=12$, $C|_G \sim \exp(2\pi i \tau_G(u))\sim u^{-1}$ and $R[C|_G]=-R[u]=-12$.

\subsubsection{Central charges}

\paragraph{6d anomalies.} 
Suppose that the 4-form appearing in \eqref{eq:-1}, now including the second Chern class $c_2(R)$ of the $\SU(2)_R$ background field, is given by
\begin{align}
I_4  = d c_2(R) + \frac14 p_1(T) + \frac14  \Tr F^2_F- \frac14  \Tr F^2_G  \label{eq:IGS}
\end{align}
The explicit value of $d$ can be determined by the method explained in \cite{Ohmori:2014kda} but it is not important here.
The contribution to the 6d anomaly polynomial from \eqref{eq:IGS}  is
\begin{align}
	\frac{1}{2}I_4^2\supset \frac{1}{4}d c_2(R)p_1(T)+\frac{1}{32}p_1(T)^2 + \frac1{16} p_1(T)\tr F_F^2.
\end{align}
Therefore, the changes in the coefficients $\alpha,\beta,\gamma,\kappa$ of \eqref{eq:coeffs} are
\begin{align}
	\delta\alpha = \frac{1}{32},\; \delta\beta=\frac{1}{4}d,\; \delta\gamma=0, 
	\; \delta\kappa=\frac{1}{16}.
	\label{eq:delal}
\end{align}

\paragraph{4d central charges.}
We now would like to determine the changes in $a$, $c$, $k$ in 4d. To do this, we use the method of \cite{Shapere:2008zf} recalled above.
 Putting the theory on a curved manifold via twisting leads to the path-integral \eqref{eq:twistedpathintegral}. 

As before, we denote by $u$ the coordinate of $\mathcal H$. 
We have one singularity at $u=0$ giving the 4d SCFT of our interest,
and there are  two  additional hypermultiplet points at $u= 1,\lambda$ where $\lambda$ is the function of the  complex moduli $\tau$ of the torus on which we compactify the 6d theory. 
We denote by $R_{0,1,\lambda}$, the R-charge of $u$ near $u=0,1, \lambda$.
Then, the measure factors $A$,$B$ and $C$ transform under $ (u-p) \to \text{exp}(iR_{p}\alpha)(u-p)$ (where $p=0,1,\lambda$) as
 \begin{equation}
 A^{\chi}B^{\sigma}C^n \to \exp[i\{(4\delta a_p -2 \delta c_p) \chi + 3\delta c_p \sigma + \delta k_p n\}\alpha] A^{\chi} B^{\sigma} C^{n} \label{eq:changeone}
 \end{equation} 
where $\delta a_p$, $\delta b_p$ and $\delta k_p$ are differences of $a$, $b$ and $k$ between the theory on $u = p$ and the theory on a generic point of $\mathcal H$. This is just the anomaly matching of the $\U(1)_R$ anomaly \eqref{eq:U1Ranom} discussed above.

Next consider very large $|u|$ region.
In this region, $\mathcal H$ looks like a cylinder $\log u \sim \phi + 2 \pi i b$.
By the dimensional reduction of \eqref{eq:-1}, the $b$ has a coupling $2\pi b I_4$. In the topologically twisted theory, 
the $I_4$ of \eqref{eq:IGS} becomes
\beq
I_4= -\frac{d}2 \chi + \frac34 (1-d ) \sigma + n_F -  n_G \label{eq:IGS2}
\eeq
where we used the fact that  
$c_2(R) = -\frac12 \chi -\frac14 p_1(T)$ due to the topological twist,
and $\sigma = p_1(T)/3$. We abuse the notation for $\chi, \sigma$ and $n$ to
mean the densities of the Euler number, signature and anti-instanton number as well as their integrals, e.g., $n=\frac{1}{4}\Tr F^2$.
Using \eqref{eq:IGS2} and noting that $2\pi i b I_4$ should be completed as $\log (u) I_4$ due to holomorphy, 
we can determine the factor $A^{\chi} B^{\sigma} C^n $ as
\beq
A^{\chi} B^{\sigma} C^n \sim \exp\left[\int \log(u) I_4 \right]=(u^{-\frac{d}2})^\chi (u^{\frac34 (1-d )})^\sigma u^{n_F} 
\eeq
and in particular, the phase shift under $u \to e^{i\alpha} u$ is given as
\begin{equation}
A^{\chi} B^{\sigma} C^n \to \text{exp}\biggl{[}i\alpha \bigl{(}-\frac{d}2 \chi + \frac34 (1-d) \sigma + n_F \bigr{)} \biggr{]} A^{\chi} B^{\sigma} C^n. \label{eq:changetwo}
\end{equation}

Now consider a circle $S^1$ going once at a large value of $|u|$. The phase shift is given by \eqref{eq:changetwo} with $\alpha=2\pi$.
Then we shrink this circle so that it becomes small circles around each of the singular points $u=0,1,\lambda$. 
The phase shift around each circle is given by \eqref{eq:changeone} with $\alpha=2\pi/R_p$.

It is known that $B$ and $C$ are single valued functions of $u$~\cite{Witten:1995gf}. 
Then the phase shift around the large circle should be the same as the sum of the phase shifts around the singular points.
First, for $C$ we get
\begin{equation}
 1 =  \sum_{u=0,1,\lambda} \frac{\delta k_u }{R_u} = \frac{\delta k_0}{R_0} \label{eq:monodromyrelation}
\end{equation}
where we  used the fact that $\delta k_{1,\lambda} =0 $ because at $u=1, \lambda$ only an additional hypermultiplet appears which is not charged under
the non-abelian flavor group $F$. 
Therefore we can determine the change in the flavor central charges:
 \begin{equation}
  \delta k= R_0=12 =192 \delta \kappa, \label{eq:delk}
  \end{equation}
where $\delta \kappa$ is given in \eqref{eq:delal}.

Next, for $B$ we get
\beq
\frac{3}{4}(1-d)= \sum_{u=0,1,\lambda} \frac{ 3 \delta c_u }{R_u}.
\eeq
The $\delta c$ at $u=1, \lambda$ comes from a free hypermultiplet and it is given as $\delta c=c_{\rm hyper}=1/12$.
The $\U(1)$ multiplet containing $u$ is IR-free at $u=1, \lambda$ and hence the R-charge is that of the free vector multiplet, $R_{1,\lambda}=2$.
Therefore we get
\beq
\delta c_0 = 2-3d= 64 \delta \alpha -12 \delta \beta -8 \delta \gamma.\label{eq:delc}
\eeq
where $\delta \alpha, \delta \beta$ and $\delta \gamma$ are given in \eqref{eq:delal}.

Finally, we consider $A$. 
In this case, $A$ is not a single valued function \cite{Witten:1995gf}.
However, the nontrivial monodromy of $A$ is fixed by the Seiberg-Witten curve of the $\U(1)$ multiplet of $u$. 
The equation \eqref{eq:E1curve} implies that the Seiberg-Witten curve is completely the same as that of rank-1 E-string theory on $T^2$.
Therefore, the ratio $A(u)/A_E(u)$ is single-valued, where $A_E(u)$ is the $A$-factor of the rank-$1$ E-string theory on $T^2$.

This  $A_E(u)$ is known to behave as $u^{1/2}$ around $u \sim \infty$ as can be seen from the analysis of the 
$E_8$ Minahan-Nemeschansky theory \cite{Shapere:2008zf} or from the fact that the study of the 6d anomaly polynomial gives $d=-1$ \cite{Ohmori:2014kda}.
Therefore,
\begin{equation}
A(u)/A_E(u) \sim u^{-(d+1)/2}. \label{eq:AA}
\end{equation} 
Furthermore, the hypermultiplet contributions cancel out in the ratio $A(u)/A_E(u)$ at $u=1, \lambda$.
Therefore \eqref{eq:AA} is actually valid over the whole ${\mathcal H}$. We get
\beq
-\frac{d+1}{2}=\frac{\delta (4a-2c)_0-R[A_E]_0}{R_0}
\eeq
where $R[A_E]_0$ is the R-charge of $A_E$ at $u=0$. It is given as $A_E(u)=(\partial u_E/\partial a_E)^{1/2}$ \cite{Shapere:2008zf}
and hence $[A_E(u)]=5$. Thus
\beq
\delta (2a-c)_0=-3d-\frac{1}{2}=-16 \delta\alpha -12 \delta\beta -28 \delta\gamma.\label{eq:del2a-c}
\eeq
Combining \eqref{eq:delk}, \eqref{eq:delc} and \eqref{eq:del2a-c} with the assumption of the induction, the proof of \eqref{eq:ack} is completed.

\subsection{Examples}\label{sec:centmatter}
\subsubsection{General-rank $E_8$ theories}
As first examples of our general analysis, let us first consider the E-string theory of general rank.
When put on $T^2$, this is known to reduce to the general-rank version of the $E_8$ theory of Minahan and Nemeschansky. 
The central charges $a$, $c$ and $k$ of these theories were found in \cite{Aharony:2007dj}: 
\begin{align}
a  &= \frac{3}{2}Q^2+\frac{5}{2}Q-\frac{1}{24}, \\
c  &= \frac{3}{2}Q^2+\frac{15}{4}Q-\frac{1}{12}, \\
k_{E_8} &= 12Q, \\
k_{\SU(2)_L} & = 6Q^2 -5Q -1,
\end{align} where $Q$ is the rank.

The anomaly polynomial of 6d higher-rank E-string theories was obtained in \cite{Ohmori:2014pca}. 
The relevant coefficients in the anomaly polynomial  are 
\begin{align}
	\alpha = \frac{7 (30Q-1)}{5760},\;\beta= \frac{-Q(6Q+5)}{48},\;\gamma=\frac{1-30Q}{1440}.
\end{align}
and 
\begin{equation}
\kappa_{E_8} = \frac{Q}{16} ,~~~ \kappa_{\SU(2)} = \frac{1}{32}Q^2  -\frac{5}{192} Q - \frac1{192}
\end{equation}
We can check that the formulas \eqref{eq:ack}  are indeed satisfied.

\subsubsection{Central charges of minimal conformal matter on $T^2$}
As second examples, let us consider the central charge of the 6d $(G,G)$ minimal conformal matter on $T^2$. The anomaly polynomial of that theory was obtained in \cite{Ohmori:2014kda}. The relevant coefficients in the anomaly polynomial are
\begin{align}
	\alpha &= \frac{7}{5760}(1+\mathrm{dim} (G)),\;\beta=\frac{1}{48}(\mathrm{dim}(G)-\chi_\Gamma  |\Gamma|),\nonumber \\ 
	\gamma &= \frac{-1}{1440}(1+\mathrm{dim}(G)),\; \kappa_G =\frac{h^\vee_G}{96}.
\end{align}
where $|\Gamma|$ is the number of elements of the discrete group $\Gamma$ used in the orbifold ${\mathbb C}^2/\Gamma$, 
and $\chi_\Gamma :=1+\rank(G) - 1/|\Gamma|$.
From \eqref{eq:ack}, we obtain the central charges as 
\begin{align}
	a= \frac{1}{24}(1+6 \chi_{\Gamma}|\Gamma|-5 \mathrm{dim}(G)),\;\;c=\frac{1}{12}(1+3 \chi_{\Gamma}|\Gamma|-2\mathrm{dim}(G)), k_G = 2h^\vee_G. \label{eq:acconformalmatter}
\end{align}

Then, we compute the central charges of the class S theory of type $G$ on a sphere with two full punctures and a simple puncture. The relevant formula \cite{Chacaltana:2012zy} is 
\begin{align}
a &= a_\text{simple} + 2a_\text{full} -\frac13 h^\vee_G \dim(G) -\frac5{24} \rank(G), \label{eq:classSa}\\
c &= c_\text{simple} + 2c_\text{full} -\frac13 h^\vee_{G} \dim(G) -\frac16 \rank(G), \label{eq:classSc}\\
k_G &= k_\text{full} \label{eq:classSk},
\end{align}
where $a_\text{simple}$ and $a_\text{full}$ are the contribution from the simple and full puncture, respectively. The contributions from the punctures are given by \cite{Chacaltana:2012zy}
\begin{align}
a_\text{simple} &= \frac1{24}(6|\Gamma|\chi_\Gamma +1 ), \; a_\text{full} = \frac1{24}(4h^\vee_G \dim(G) -\frac52 \dim(G) +\frac52 \rank(G)), \nonumber \\
c_\text{simple} &= \frac1{12}(3|\Gamma|\chi_\Gamma + 1), \; c_\text{full} = \frac1{12}(2h^\vee_G \dim(G) -\dim(G) +\rank(G)), \nonumber \\
k_\text{full}  &=  2h^\vee_G. \nonumber
\end{align}
Substituting these equations into \eqref{eq:classSa}, \eqref{eq:classSc} and \eqref{eq:classSk}, we obtain the same central charges as \eqref{eq:acconformalmatter}. This provides a non-trivial check both for the central charge formula in \eqref{eq:ack} and the duality between the minimal conformal matter on $T^2$ and the class S theory.

\section{Conclusions and discussions}\label{sec:conclusions}
In this paper we found that the world volume theory of a single M5-brane on the tip of an ALE space of type $G=A,D,E$,
namely the 6d $(G,G)$ minimal conformal matter, gives a type $G$ class S theory with a sphere accompanied by two full-punctures and a simple puncture, namely 4d generalized bifundamental, by means of  $T^2$ compactification.

We have given several evidences on this statement.
We provided  the matching of coulomb branch dimensions and the Higgs branch geometry, and we checked the agreement of the Seiberg-Witten curve in the case of type $D$ in a certain corner of the moduli space, by exhibiting the   ``base-fiber duality'' indicated by the 6d brane construction at the level of the 4d Seiberg-Witten curves.
We also developed a new method to study the central charges of the $T^2$ compactification of a class of the 6d SCFTs that we call \epf, and applied this technique to the minimal conformal matters. We again found agreement with the central charges of the class S theories. 
With these checks, we find that our proposed identification is well established. 

Let us discuss some of the future directions.
\paragraph{Other \epf theories}
There are many \epf theories which are neither $(G,G)$ minimal conformal matters nor higher-rank E-string theories.
For $T^2$ compactifications of all of those, we showed that the formula \eqref{eq:ack} holds.

Some of these theories can be obtained by considering ``fractional M5-branes'' on ALE singularities:
\begin{itemize}
\item The $(E_7,\SO(7))$ minimal conformal matter, namely a ``half M5-brane'' on top of $E_7$ singularity,
\item the $(E_8,G_2)$ minimal conformal matter which is a ``third M5-brane'' on $E_8$ singularity, 
\item and the $(E_8,F_4)$ minimal conformal matter which is a ``half M5-brane'' on $E_8$ singularity.
\end{itemize}

For the $(E_7,\SO(7))$ minimal conformal matter, we can find a candidate of the corresponding 4d theory in the list of $E_6$ tinkertoys \cite{Chacaltana:2014jba}.
Conbining the method of \cite{Ohmori:2014kda} and the formula \eqref{eq:ack}, we find the central charges of $T^2$ compactified $(E_7,\SO(7))$ minimal conformal matters are 
\begin{align}
	a=\frac{119}{8},\; c=\frac{35}{2},\; k_{E_7}=24,\; k_{\SO(7)}=16.
\end{align}
Those numbers are exactly the same as the conformal central charges of $E_6$ fixture with punctures $E_6(a_1)$, $2A_1$ and the full puncture, where the notation of the punctures are of \cite{Chacaltana:2014jba}. 

Similarly, the candidates for the $(E_8,G_2)$ and $(E_8,F_4)$ minimal conformal matter might be found in $E_7$ or $E_8$ fixtures. But the list of $E_7$ and $E_8$ fixtures are not yet available. 

Another natural series of \epf theories can be found by considering theories on M5-branes on the intersection of an end-of-the-world brane and an ALE singularity locus.
In contrast to the minimal conformal matters, the theories are endpoint-trivial for all integer numbers of $M5$-branes, and therefore there are infinitely many of them.
It would be interesting to search 4d corresponding theories in known 4d SCFTs.

\paragraph{Non \epf theories}
The worldvolume theories on multiple coincident M5-branes on an ALE singularity locus, are not \epf.  Thus the approach of this paper cannot be directly applied and new methods need be introduced to investigate such theories.

The \Nequals{(1,0)} SCFTs which are defined by the F-theory with Hirzebruch's surface $F_n$
as its base are other cases recently studied in \cite{Haghighat:2014vxa}.
Although the structure of the base $F_n$ is very straightforward, in that it contains just one $-n$ curve,
our method cannot be applied to these when $n\ge 3$. 
It would be interesting to devise a method that can be applied to the $T^2$ compactification of any 6d SCFT.

\paragraph{Compactification with general Riemann surfaces and punctures}
Our ultimate goal would be  to study compactifications of 6d \Nequals{(1,0)} theories with general Riemann surfaces with punctures giving 4d \Nequals1 theories rather than 4d \Nequals2.
Although there clealy is a \Nequals1 theory defined by compactification of a \Nequals{(1,0)} theory with a genus $g\ge 2$ Riemann surface, we do not have any tools to identify or investigate such theory.
In contrast to the \Nequals{(2,0)} case, the theory on the tube is already non-trivial,  preventing us from studying on S-dualities between compactified theories. 
The $T^2$ compactified theories studied in this paper might be a clue to find out the tube theories if one can find an appropriate boundary conditions at the ends of the tube.


The authors hope to come back to these questions in the future. 

\section*{Acknowledgments}
KO and HS are partially supported by the Programs for Leading Graduate Schools, MEXT, Japan,
via the Advanced Leading Graduate Course for Photon Science
and via the  Leading Graduate Course for Frontiers of Mathematical Sciences and Physics, respectively. 
KO is also supported by JSPS Research Fellowship for Young Scientists.
YT is  supported in part by JSPS Grant-in-Aid for Scientific Research No. 25870159,
and in part by WPI Initiative, MEXT, Japan at IPMU, the University of Tokyo.
The work of KY is supported in part by DOE Grant No. DE-SC0009988.
 
\appendix
 
\bibliographystyle{ytphys}
\bibliography{ref}

\end{document}